\documentstyle[preprint,aps,prl]{revtex}
\tightenlines

\newcommand{\beq}{\begin{equation}}
\newcommand{\eeq}{\end{equation}}
\newcommand{\beqa}{\begin{eqnarray}}
\newcommand{\eeqa}{\end{eqnarray}}
\newcommand{\eps}{\epsilon}

\begin{document}
\draft
\title{Stripes and superconductivity in one-dimensional
 self-consistent model}
\author{S.I. Matveenko}
\address{Landau Institute for Theoretical Physics, Kosygina Str. 2, \\
119334, Moscow, Russia\\}

\date{\today}
\maketitle
\begin{abstract}
We show that many observable properties of high temperature superconductors
can be obtained in the frameworks of one-dimensional
self-consistent model
with included superconducting correlations.
Analytical solutions for spin, charge and superconductivity order parameters
are found.
  The ground state  of the model at low hole doping
is a spin-charge  solitonic  superstructure.
Increasing of doping leads to the phase transition to superconducting
phase. There is a region of doping where superconductivity,
spin density wave and charged stripe structure coexist.
 The charge density modulation presents  in the vicinity of vortices
(kinks in the 1D model) in the superconducting state.

\end{abstract}

\pacs{PACS numbers: 71.10.Fd, 74.72.-h, 64.60.-i, 71.27.+a}

Recently discovered stripe phases in doped antiferromagnets (cuprates and
nickelates) \cite{tranq} have attracted attention to the problem of
 coupled spin and charge order parameters in the electron systems.
Theoretical \cite{zaan,mach,schul,mar} and experimental
\cite{exp1,exp2,exp3,exp4,exp5} evidence indicate  the possibility
that their ground state exhibits spin and charge density waves
(SDW and CDW), either competing, or coexisting with
superconductivity. Numerical mean-field calculations
\cite{zaan,mach,schul} suggest a universality of the spin-charge
multi-mode coupling phenomenon in repulsive electronic systems of
different dimensionalities.
 Families of the cuprate high-transition-temperature
superconductors show antiferromagnetism and
superconductivity. For the $La_{2-x}Sr_xCuO_4$  family there
 is another ordering tendency - unidirectional
charge-spin density wave, i.e. ``stripes''.
Recent neutron scattering experiment of Lake {\it et al}
shows that  moderate magnetic field makes fluctuating stripes quasi
static\cite{exp1}.
An important development in the theory of the cuprate superconductors
is the prediction that in addition to
antiferromagnetism and superconductivity there is
 a tendency toward
stripe ordering\cite{zaan,mach,schul}.
 This prediction is corroborated by experiments\cite{tranq,ll}.
A recent neutron scattering experiment shows that a moderate magnetic field can
turn a fluctuating stripe order into
a quasi static one in the optimum doped cuprates.\cite{exp2}
The vortex
state can be regarded as an inhomogeneous mixture of a superconducting
spin fluid and a material containing a nearly ordered antiferromagnet.

In this paper we present the one-dimensional effective model
describing stripe phase at low hole doping and superconductivity state at
higher doping.
An exact analytical solution of the Hartree-Fock problem
 at and away from half-filling is found.
This solution provides a unique possibility to investigate  analytically the
structure of the periodic  spin-charge solitonic superlattice.
It also demonstrates fundamental importance of the higher order
commensurability effects, which result in special stability points along the axis of
concentrations of the doped holes.
Our  theory predicts an amazing duality between the spin density wave and
superconducting order,
and implies the presence of stripes near a
superconducting vortex, and superconductivity near a stripe dislocation.
 Though there is no long-range order in the purely
one-dimensional system due to destructive influence of fluctuations, real cuprates
are three-dimensional, and therefore, the long-range order survives in the ground
state.
Hence, we believe,
that one-dimensional mean-field solutions contain universal features
of the  stripe phase, which are stabilized in higher dimensions.
Single-chain analytical solutions may be used
 as building blocks for the stripe
and superconducting phase
in quasi
two(three)-dimensional system of parallel chains.

The Hamiltonian $H = H_0 + H_s$ consists of two parts:
the Hubbard Hamiltonian with  on-site repulsion
$U > 0$

\begin{equation}
H_0 =\displaystyle -t\sum_{\langle i,j\rangle
\sigma}c^{\dagger}_{i,\sigma}c_{j,\sigma}+ U\displaystyle\sum_{i}
\hat{n}_{i,\uparrow} \hat{n}_{i, \downarrow}
- \mu \sum_{i. \sigma} \hat{n}_{i, \sigma} ,
  \label{hubbard}
\end{equation}
and the interaction part including superconducting correlations
\beq
H_s = \sum_{i} \Delta_s(i)c^{\dagger}_{i,\uparrow}c^{\dagger}_{i,\downarrow} + h.c.,
\label{hs}
\eeq
where $\Delta_s$ is the superconducting order parameter, $\mu$ is the chemical
potential.
The case  of the Hubbard model (\ref{hubbard})
was considered in details earlier \cite{mm}.
The self-consistent analytical solution  for the charge-spin
solitonic superstructure
was found as a function of a hole doping.
It was shown that effects of commensurability
 led to a pinning of stripe structure at rational filling
points $|\rho - 1| = m/n$.

In the continual self-consistent approximation the effective Hamiltonian can
be derived similar \cite{mm}. We obtain
\begin{eqnarray}
H = \int dx \{ {\bf\Psi}_{\sigma}^{\dag}\left(-i\frac{\partial}{\partial
x}\right)\hat{\sigma}_z {\bf\Psi}_{\sigma} + \Delta(x) {\bf\Psi}_{\sigma}^{\dag}
\hat{\sigma}_+ {\bf \Psi}_{\sigma}  + \nonumber \\
\Delta^*(x){\bf\Psi}_{\sigma}^{\dag}
 \hat{\sigma}_- {\bf \Psi}_{\sigma} +
 \alpha \rho(x){\bf \Psi}_{\sigma}^{\dag} {\bf \Psi}_{\sigma} +\quad \nonumber \\
\Delta_s ( -\Psi^{\dag}_{+,\uparrow}\Psi^{\dag}_{-,\downarrow} +
\Psi^{\dag}_{-,\uparrow}\Psi^{\dag}_{+,\downarrow} ) +\qquad \nonumber \\
\Delta^*_s ( -\Psi_{-,\downarrow}\Psi_{+,\uparrow} +
\Psi_{+,\downarrow}\Psi_{-,\uparrow} ) +\qquad \nonumber \\
\frac{|\Delta |^2}{\pi\lambda} +\frac{|\Delta |_s^2}{\pi\lambda_s}
-\frac{\alpha}{2}\rho^2 \},\qquad
\label{heff}
\end{eqnarray}

\noindent where $\lambda = 2\alpha /\pi$ is a dimensionless spin coupling
constant, $\lambda_s$ is a  dimensionless superconductor coupling constant,
 $\hat\sigma_{z,x}$ are the Pauli matrices,
$2\hat\sigma_{\pm}=\hat\sigma_x \pm \hat\sigma_y$,
$\alpha = U/4t $; the Plank constant is taken as
unity, and the length is measured in the units of the lattice (chain) period $a$.
 In
these units momentum and wavevector are dimensionless, and velocity and energy
posses one and the same dimensionality.
The vector ${\bf{\Psi_\sigma}}^{T}\equiv (\Psi_{\sigma +},\Psi_{\sigma -})$ is
defined
in terms of the right- left-moving $\Psi_{\sigma\pm} $, which constitute the wave
function:

\beq
\Psi_{\sigma}(x)= \Psi_{+,\sigma} e^{ik_F x} +
\sigma \Psi_{-,\sigma} e^{-ik_F x}
\label{psi}
\eeq

where $\sigma =\pm 1$ for a spin $\uparrow$ and $\downarrow$ respectively. The
Fermi-momentum is $k_F= \pi\bar\rho/2$, where
in the case of half-filling the
average number of electrons per site equals $\bar\rho = 1$, i.e. $k_F=\pi/2$.
The slowly varying real functions $\Delta(x)$ and $\rho (x)$ are defined as
$ \langle \hat{n}(x) \rangle =   \rho (x)$,
$ \langle \hat{S}^{z}(x)\rangle = -\Delta(x)\cos(\pi x)/\alpha$.
The continual approximation requires that $\alpha, \lambda, \lambda_s
\ll1$ (weak coupling limit). Note that the constraint $\lambda = 2\alpha
/\pi$ for the Hubbard model is not necessary in a general case, our results
remain valid for independent $\alpha, \lambda, \lambda_s
\ll1$.

Introduce $\bar \rho $ and $\tilde \rho $ as $\rho (x) = \bar \rho
+ \tilde \rho (x)$, $\int\tilde\rho (x) dx = 0$. Then the term
$\alpha \bar\rho \Psi^{\dag}\Psi$ in Eq. (\ref{heff}) is the shift
of the chemical potential or the energy, and the term $\alpha
\tilde\rho \Psi^{\dag}\Psi$ can be excluded by the unitary
transformation (see \cite{mm,mu})
\[
\Psi_{\pm }(x) \longrightarrow \exp(\mp i\alpha \int^{x} \tilde \rho \, dx' )
\Psi_{\pm}(x).
\]
Under this transformation the spin order parameter modifies as
$\Delta (x)  \longrightarrow \exp(2 i\alpha \int^{x} \tilde \rho \, dx' )
\Delta(x)$.

We can diagonalize the total Hamiltonian $H = H_0 + H_s$ by performing a
 unitary Bogolubov transformations

\beq
\Psi_{\sigma}(x) = \sum_n \gamma_{n, \sigma} u_{n, \sigma}(x)-\sigma
\gamma^+_{n, -\sigma} v^*_{n, -\sigma}(x)
\label{tr}
\eeq
which have the form
\beq
\Psi_{\pm,\sigma} = \sum_n \gamma_{n,\sigma} u_{\pm}
\pm \gamma_{n,-\sigma}^+ v_{\mp}^*
\label{gamma}
\eeq
\noindent in terms of right and left
 components $u_{\pm}$, $v_{\pm}$ defined as

\beq
f_{\sigma}(x)= f_{+,\sigma} e^{ik_F x} +
\sigma f_{-,\sigma} e^{-ik_F x}, \quad f=u,v.
\eeq

New operators $\gamma$, $\gamma^+$ satisfy the fermionic commutative
relations $
\{\gamma_{n, \sigma}, \gamma^+_{m, \sigma^{\prime}}\} = \delta_{m,n}
\delta_{\sigma, \sigma^{\prime}}$.
The transformations (\ref{tr}) must diagonalize the Hamiltonian $H$:
\beq
H =  E_g + \sum_{\eps_n >0} \eps_n \gamma_{n,\sigma}^+ \gamma_{n,\sigma},
\label{eg}
\eeq
\[
E_g = \sum_{\eps_n <0} \eps_n + \int dx \left(
\frac{|\Delta |^2}{\pi\lambda} +\frac{|\Delta |_s^2}{\pi\lambda_s}
-\frac{\alpha}{2}\rho^2 \right),
\]
where $E_g$ is the ground state energy and $\eps_n >0$ is the energy
of excitation $n$.

Following \cite{bdg} we obtain the eigenvalue equations
\beq
\hat{H}\chi = \eps \chi,
\label{H}
\eeq
where
\[
\hat{H} =
\left(
\matrix{-i\frac{\partial}{\partial x} + \alpha \rho & \Delta & \Delta_s &0 \cr
     \Delta^*  & i\frac{\partial}{\partial x} + \alpha \rho & 0& \Delta_s \cr
\Delta_s^* &0 & i\frac{\partial}{\partial x} - \alpha \rho & \Delta \cr
   0 & \Delta_s^* & \Delta^* & -i\frac{\partial}{\partial x} - \alpha \rho }
\right),
\]
$\chi^T = (u_+, u_-, v_+, v_- )$,
and self-consistent conditions

\beq
\rho(x) = 2\sum [(u_+^* u_+ + u_-^* u_- )f +
 (v_+^* v_+ + v_-^* v_- )(1-f)]
\label{s1}
\eeq

\beq \Delta(x) = -4\lambda [\sum u_-^* u_+ f - \sum v_-^* v_+
(1-f)] \label{s2} \eeq

\beq
\Delta_s = 2 \lambda_s \sum (1-2f) [v_+^* u_+ + v_-^* u_- ],
\label{s3}
\eeq
where $f = {1}/({\exp[\eps /T] +1})$. We omitted spin indices since in our
representation for wave functions all equations are diagonal over spin.

At first, consider homogeneous state with
constant $\Delta =|\Delta| \exp [i\varphi ] $,
 $\Delta_s = |\Delta_s | \exp [i\varphi_s ]$
and $\rho(x) = \bar{\rho} \equiv N/L$.
The  average spin density has the form
 $<S_z > \propto Re (\Delta \exp (2i k_F x)$.
Neglecting trivial dependence on $\bar{\rho}$
we obtain two branch spectrum
\beq
\eps^2_{\pm}= k^2 + (|\Delta | \pm |\Delta_s|^2 )^2,
\eeq
with wave functions $u, v \propto \exp [ikx]$ satisfying the symmetry
relations
\beq
v_+ =  \pm u_- \exp i[\varphi -  \varphi_s],\,
v_- =  \pm u_+ \exp -i[\varphi +  \varphi_s].
\label{sym}
\eeq
The self-consistent equations read

\beq
|\Delta| =\frac{\lambda}{L} [F_+ + F_- ], \quad |\Delta_s | =
\frac{\lambda_s }{L}  [F_+ - F_- ],
 \eeq
 where $F_{\pm} =
\sum_{\eps} [(|\Delta| \pm |\Delta_s |)/\eps_{\pm}] \tanh
[\eps_{\pm}/T]$. At zero temperature we obtain \beq
\frac{2}{\lambda} = \log \frac{4\eps_F^2}{||\Delta|^2 -
|\Delta_s|^2 |} + \frac{|\Delta_s|}{|\Delta|} \log \left|
\frac{|\Delta| - |\Delta_s|}{|\Delta| + |\Delta_s|} \right|
 \eeq
The second equations is derived from the first one by substitution
$\lambda \rightarrow \lambda_s$, $\Delta \leftrightarrow
\Delta_s$. The minimum of the ground state energy $E_g$ is
achieved at the state $\Delta = 2\eps_F \exp [-1/\lambda ]$,
$\Delta_s = 0$ for the case $\lambda > \lambda_s$, and $\Delta_s =
2\eps_F \exp [-1/\lambda_s ]$, $\Delta = 0$ for the case $\lambda
< \lambda_s$

In general case parameters $\lambda$, $\lambda_s$ depend on the
doping concentration $h=|\rho - 1| $. It is well known that the
coupling constant $\lambda$ monotonically decreases
 with doping from $\lambda_0$ at $\rho = 1$ to the value $\lambda_0 /2$ in the
limit $|\rho - 1| \gg \Delta /v_F$ (due to the absence of umklapp scattering
 at $ \rho \neq 1$) \cite{br}.
If we suppose that superconducting part $H_s$ comes from next neighboring
site repulsion (as considered for 2D $Cu O$ plane model)
$H_s \sim V \rho_n \rho_{n \pm 1}$,
then the self-consistent equation becomes $\Delta_s \sim V \langle
\Psi_{n,\downarrow} \Psi_{n \pm 1 , \uparrow}\rangle \to 2V \cos k_F a
\langle
\Psi_{\downarrow}(x) \Psi_{\uparrow}(x)\rangle$ in the continual approximation.
The coupling constant $\lambda_s \sim \frac{2}{\pi} V \cos\frac{\pi \rho }{2}$
increases with hole doping $h = 1- \rho$.
If the ground state of undoped system is antiferromagnet state
 ($\lambda > \lambda_s$), phase transition to superconducting state
will take place at some doping $h_c$ where $\lambda = \lambda_s$.
Two phases (SDW and SC) with $\Delta = \Delta_s \neq 0$
 can coexist at this point. For detailed investigation of
the phase transition more rigorous consideration of quantum fluctuations
is necessary.

So far we considered uniform state with $\Delta$, $\Delta_s = const$.
 Since  symmetry relations between wave function
components (\ref{sym}) are independent of absolute values
 ($ |\Delta|$, $|\Delta_s| $),
we assume that these relations are valid also in a general case of
nonuniform order parameters. Substituting (\ref{sym}) to (\ref{H}) we
obtain in the case of constant phases $\varphi$, $\varphi_s$
equations

\beq
[-i \sigma_z \frac{d}{dx} + \tilde{\Delta}\sigma_+ + \tilde{\Delta}^*
\sigma_- ] {\bf u} = \eps {\bf u },
\label{eq1}
\eeq
where ${\bf u}^T = \{ u_+, u_- \}$,
$\tilde{\Delta} = (\Delta \pm \Delta_s )\exp [i\varphi ]$.
These equations are eigenvalue equations for the Peierls model,
were studied in \cite{br,db}.
The dependance  on $\rho$  in (\ref{eq1}) was
 excluded by  means of wave function transformation
$u_{\pm},v_{\pm} \to \exp\{\mp i \alpha\int \rho dx \} u_{\pm},v_{\pm}$.
The term $ \alpha \int dx \rho^2 /2$ in the  total energy $E_g$
is responsible for commensurate effects and pinning of the system
at rational doping ($h = m/n$) points \cite{mm}.

 Consider the system  with $\lambda
> \lambda_s$.  At $\bar{\rho} = 1$ the ground state is antiferromagnet with constant
$\Delta = \Delta_0 $,
 $\rho= \bar{\rho}$ and $\Delta_s = 0$.
As a result of doping kink states are formed with local level $\eps = 0$ at
the center  of the gap $2\Delta$.
The  single kink solution  of (\ref{eq1}) is
$\tilde{\Delta}_1 = \Delta + \Delta_s = \Delta_0 \tanh (\Delta_0 x + a/2)$,
 $\tilde{\Delta}_2 = \Delta - \Delta_s = \Delta_0 \tanh ({\Delta}_0 x - a/2)$
with arbitrary shift $a$.
The wave functions and the excitation spectrum read
\beq
u_{\pm} \propto (\pm \eps  + k +
i\Delta_0 \tanh \xi )e^{ikx} e^{\pm i\pi/4},\quad \eps^2 = k^2 + \Delta_0^2,
\eeq
\beq
u_{0,\pm} \propto \frac{\exp[\pm i \pi /4]}{\cosh^2 \xi}, \quad \eps = 0,
\eeq
where $\xi = \Delta_0 x \pm a/2$.
 The order parameters $\Delta$, $\Delta_s$
take form
\[
\Delta_s = \frac{\Delta_0 \sinh a}{2(\cosh^2 \Delta_0 x + \sinh^2 a/2 )},
\quad
\Delta = { \frac{\Delta_0 \tanh \Delta_0 x }{1 + \frac{\sinh^2 a/2}
{\cosh^2 \Delta_0 x }}}.
\]
For the case $a = 0$ we obtain $\Delta_s \equiv 0$, $\Delta = \Delta_0 \tanh
\Delta_0 x$, $\rho \propto 1/\cosh^2 \Delta_0 x$. It is a one stripe
solution found in \cite{mm}.
The shift $0< a \ll 1$ leads to the appearance of the region around the stripe
 with $\Delta_s \neq 0$, so that $\rho \propto 1/\cosh^2 \Delta_0 x$,
 $\Delta_s \propto a/\cosh^2 \Delta_0 x$.
The quasiparticle spectrum is independent of $a$, therefore the
equilibrium position $a$ is defined by minimization of the
potential energy
\beq
\delta W(a) = \frac{\Delta_0}{\pi}\left|
\frac{1}{\lambda_s} - \frac{1}{\lambda} \right| \frac{a}{\tanh a}
+ \frac{ \Delta_0 \alpha }{4} \frac{\frac{a}{\tanh a} - 1}{\sinh^2
a}.
 \label{wa}
 \eeq
The minimum of the energy (\ref{wa}) is reached at $a = 0$ for
\[
\gamma \equiv   \frac{\alpha \pi \lambda \lambda_s}{4 |\lambda -
\lambda_s |} - 2.5 < 0.
\]
For small $a$ the inequality $\gamma < 0$ is possible if $\lambda$
and $\lambda_s$ are not very close to each other ($|\lambda -
\lambda_s| \gtrsim \alpha \lambda \lambda_s$). The nontrivial
minimum $a \neq 0$ exist only in the small region $|\lambda -
\lambda_s | \lesssim \alpha \lambda \lambda_s$ around the
transition point $\lambda = \lambda_s$, where $\gamma >0$. Stripe
and superconductivity phases coexist in this region: $\Delta_s,
\Delta , \rho(x) \neq 0$. The equilibrium shift $a$ is small $a
\propto \sqrt{\gamma}$ if $\gamma \ll 1$, but it l logarithmically
diverges $a \propto \log \gamma $ in the limit $\lambda_s \to
\lambda $.

So we obtain that an increasing of doping for the system with
$\lambda
> \lambda_s$, $\gamma < 0$ at $\rho = 1$ leads to the forming of
the periodic structure of charged kinks $\Delta = \Delta_0 \tanh
\Delta_0 x$, which acquires  the form at $h = |\rho - 1| \gg v_F
/\Delta_0$ \cite{mm}
\[
{\Delta} \sim \Delta_0 \sqrt{k} {\rm sn}
 [\Delta_0 x  /\sqrt{k}, k],\,
\rho (x) - \bar{\rho} \propto \Delta^2 - \bar{\Delta}^2 ,
\]
where $K(k)$ is  the Elliptic Integral of the first kind,
 $\rm{sn}(.,k)$ is the Jacobi elliptic function,
$|\bar{\rho} - 1| = \Delta_0 /(2 K(k) \sqrt{k})$.

The superconducting order parameter order $\Delta_s \neq 0$
appears in the considered case at some higher doping level where
$\gamma$ becomes positive. In a small region $|\lambda - \lambda_s
| \lesssim \alpha \lambda \lambda_s$ around the transition point
$\lambda = \lambda_s$, where $\gamma > 0$, superconductivity and
spin/charge orders coexist: $\Delta_s, \Delta , \rho(x) \neq 0$.
In the particular case of the model with $\alpha = 0$ this region
is reduced to the point $h = h_c$. A more complicated analysis
beyond the scope of the used mean field approximation is required
at this point to take into account strong quantum fluctuations,
including the zero mode due to the degeneration of the ground
state with respect to the shift $a$ of two sublattices.

The   opposite region $\lambda < \lambda_s$ can be studied using
the duality properties of the model.
 It is easy to see that
eigenvalues $\eps_n$  of equations  (\ref{H}) are invariant under
transformation $\Delta \longleftrightarrow \Delta_s$. Indeed,
if we simultaneously exchange $\Delta \longleftrightarrow \Delta_s$
and $ u_- (x) \longleftrightarrow v_+ (x)$ in Eq. (\ref{H})
the Hamiltonian (\ref{H}) is not changed
(without unimportant terms with $\rho(x)$) .
Therefore the ground state energy $E_g$ in (\ref{eg}) is invariant under
the transformations
\[
\Delta \longleftrightarrow \Delta_s, \quad
 \lambda \longleftrightarrow \lambda_s.
\]
 Therefore we can apply obtained above  solutions
for  the superconductivity phase.
We find that in the region $\gamma > 0$
the ground state is superconductor with
$\Delta_s= const $, $\Delta = 0$.
The  one-dimensional analogue of the vortex in two- or three-dimensional
systems is
the kink:
$\Delta_s = \Delta_0 \tanh \Delta_0 x$.

Due to the duality symmetry the charge density $\rho(x)$ has the
same expression as for the kink in spin density wave. Therefore
we obtain that the charge density is not zero in the vicinity
 of the kink
\beq
\rho (x) \sim \frac{1}{\cosh^2 \Delta_0 x} \cos (2\pi |\rho - 1|).
\eeq

Similar to the previous case ($\lambda > \lambda_s $) we find that
near the transition point ($\lambda \sim \lambda_s, \,\gamma >0$)
a stripe structure can arise. In the limit $0<\gamma \ll 1$,
$|\bar{\rho} - 1| \gg \Delta_0 /v_F$ we obtain
\[
  \Delta(x) \propto \sin \pi |\bar{\rho} - 1|x,\quad
\tilde{\rho}(x) \propto \cos 2 \pi |\bar{\rho} - 1|x,\quad
a\propto \sqrt{\gamma}.
\]
To conclude,  we have found the self-consistent mean-field
analytical solution for the ground state structure of the
quasi-one-dimensional electronic system with spin, charge and
superconducting correlations.
  We have found that for an appropriate choice of
parameters the ground state is striped charge/spin density wave
structure at low hole doping. The stripe configuration is pinned
at rational points $|\rho - 1| = m/n$ with the pinning energy
$\propto \exp (- C n)$ substantial for small $n$, which can lead
to the stability of the stripe picture. The phase transition to
the superconductivity state takes place at some doping level.
  Both superconductivity and spin/charge density wave  order
parameters
 can coexist in a some  small region near this point.

The model is self-dual: The eigenfunction equations are invariant
with respect to transformations $\lambda \leftrightarrow
\lambda_s$, $\Delta \leftrightarrow \Delta_s$. Therefore
properties of superconducting state can be derived  from the low
doping consideration. In particular, we obtained that charge
stripes can exist as in low doping spin density wave state as in
superconducting state in the vicinity of spatially nonuniform
configurations of $\Delta_s$, for example, vertices (kinks in one
dimension).

Though this one-dimensional model can be applied rather to
quasi-one-dimensional systems than  to high-temperature
quasi-two-dimensional anisotropic superconductors, it shows  some
properties peculiar to high-temperature superconductors
(one-dimensional stripe structure at low doping and
superconductivity at a higher doping, etc.). Therefore our results
can be useful for understanding of high-temperature phenomenon.
For describing anisotropic properties of real systems a
two-dimensional model consideration is required to take into
account an important contribution from nodal quasiparticles.

I thank A.V. Balatsky, I. Martin, S.I. Mukhin and J. Zaanen for
stimulated discussions and K. Machida for a pointing relevant
references out. I The work was supported by RFBR
 grant No. 02-02-16354.


\end{document}